# Self-Organized Networks: Darwinian Evolution of Myosin-1

J. C. Phillips

Rutgers Univ., Piscataway, N. J.,

Abstract

Cytoskeletons are self-organized networks based on polymerized proteins: actin, tubulin, and driven by motor proteins, such as myosin, kinesin and dynein. Their positive Darwinian evolution enables them to approach optimized functionality (self-organized criticality). The principal features of the eukaryotic evolution of the cytoskeleton motor protein myosin-1 parallel those of actin and tubulin, but also show striking differences connected to its dynamical function. Optimized (long) hydropathic waves characterize the molecular level Darwinian evolution towards optimized functionality (self-organized criticality). The N-terminal and central domains of myosin-1 have evolved in eukaryotes at different rates, with the central domain hydropathic extrema being optimally active in humans. A test shows that hydropathic scaling can yield accuracies of better than 1% near optimized functionality. Evolution towards synchronized level extrema is connected to a special function of Mys-1 in humans involving Golgi complexes.

**Introduction**

The cytoskeleton gives a cell its shape, supports it, and facilitates movement. In an earlier article [1], we discussed actin, a medium-size protein with 377 amino acids, which polymerizes to form cytoskeletal filaments; actin is the most abundant protein in most eukaryotic cells. Tubulin (~ 440 amino acids), a cytoskeletal second component, is a heterodimer which polymerizes to form cylinders supported by actin [2]. Tubulin assembles to form microtubules, which act as "railways' for motor-driven intracellular transport [3,4]. A third cytoskeletal component, myosin-1, is a molecular motor, which contains nearly 2000 amino acid side groups necessarily involving long range interactions.

The evolution of molecular motors was key to the origin and diversification of the eukaryotic cell. There are three major superfamilies of motor proteins: kinesins, dyneins, and myosins. The first two act as motors on microtubule filaments, while myosins function on monomeric actin [5], and their evolution should be simpler. Myosins bind to filamentous actin and produce physical forces by hydrolyzing ATP and converting chemical energy into mechanical force [6-8].

The Darwinian evolution of myosin-1 (the simplest and oldest member of the myosin superfamily, ~ 1940 amino acids) discussed here involves the thermodynamic scaling methods used previously for actin [1]. One uses evolution itself to identify dominant hydropathic strain field waves. The method uses sliding windows of width W as an alternative to single site (W = 1) phylogenetic methods to identify long-range interactions associated with domains. One can treat thermodynamic first-order effects with the standard KD scale based on air-water enthalpy unfolding differences [9]. If the function mainly involves long-range (allosteric) second-order interactions, these can be quantified as well, using the MZ scale based on identification of self-organized criticality [10] in the solvent-accessible surface areas (SASA) of > 5000 protein amino acid segments from the modern Protein Data Base [11]. Here we will see clear distinctions between short- and long-range interactions.

**Cytoskeletons have a natural length $W^* \sim 25$**

While they differ in many details, motor proteins contain motor heads and lever arms anchored to thick filaments [8]. Uniprot P12882 identifies the myosin-1 motor sites 86 -782, and the coiled coil lever sites 843-1939. The C terminal rod-like tail sequence is highly repetitive, showing cycles of a 28-residue repeat pattern. Our key parameter is an optimized sliding window width $W^*$, over which hydropathicity values $\Psi(aa)$ are best averaged; this average is denoted by the matrix $\Psi(aa,W^*)$ [1,2]. For myosin the optimized sliding window length $W^*$ should be close to 28. At the same time, actin is a key part of the cytoskeleton, and supports myosin, with $W^* = 21$ (typical for membrane proteins) [1]. In unpublished work on tubulin, it was found that a compromise value $W^* = 25$ gave good results for the evolution of tubulin, and that value is used here too. It is not only much larger than single site matching in phylogenetics (W = 1), but also much larger than typical alpha helix or beta strand secondary segments spanning ~ 5-7 sites.

It is reasonable to suppose that evolution has refined weaker long range interactions, as the stronger short-range interactions are likely to be stable for even early proteins. Evolution of a protein can be quantified through the ratio V[ Ψ(aa,W*,later)]/V[Ψ(aa,W*,earlier)]. Here V[f] is the variance of a function f, variance being the mean of the square minus the square of the mean. The most important general feature of a globular shape is its hydropathic compactness or roughness, which is determined by its hydropathic curvatures near its core or surface extremities. The average curvatures are related to the variances [2]. Our earlier work has shown how the extremes of Ψ(aa,W*) are sensitive to conformational changes between a protein's resting state and its functional state, which are separated by small energies that are much less than unfolding energies [12,13].

**Results**

Like actin and tubulin, N-terminal myosin-1 is well conserved, with human-chicken identities ~ 91%, positives ~ 96%, and 0.1% gaps. The simplest eukaryote is single-cell fission yeast [2], with respective human similarities of 42%, 59%, and 7%. Similarly, fruit fly's similarities are 58%. 73% and 1%. The feature that has evolved the most is the gaps. Looking at the BLAST human-yeast arrays we are struck by human gap lengths of 12 (twice) and several 5 or 6 gaps. Curiously, 12 ~ 25/2, and of course 6 = 12/2, which suggests that the gaps (or insertions) strengthen W = 25 water waves. The two 12 amino acid subharmonic gaps are located near the N terminal.

According to Uniprot, structural homology rules separate a fusion region 33-82 from the motor region 86-782, while Fig. 1 shows a strong hydrophobic pivot in Ψ(aa,W*) centered near 130. The first 12 aa gap falls in the 33-82 fusion region, while the second human 12 aa gap deletes a deep hydrophilic yeast hinge. Perhaps most striking in Fig. 1 are the level pivots (hydrophobic extrema, ~ 120 and 350) and level hinges (hydrophilic extrema,150, 300, 410) found in single cell yeast but not in multicell human.

Level sets describe (see Secs. 7, 9, 10 of [13], as well as [14]) the continuum hydrodynamics of optimized curved front propagation. More simply, protein motion is optimized with level sets, because then all extremal domain motion is synchronized at similar rates, with no laggards or bottlenecks. For myocin-1 in the N-terminal region 1-420 shown in Fig. 1, this simple picture works best for monocellular yeast. Multicellular motion is facilitated and stabilized by crowded intercellular waves, so evolution may have deemed the deep yeast hinge at gap II in Fig. 1 to be unnecessary, and deleted it in all subsequent species.

Profiles for the central region 300-680 are shown in Fig. 2. As described in the Figure caption, the differences between Human and Yeast seen in Fig. 1 for the N terminal region 1-420 are reversed in the central region 300-680. It appears that evolution nearly perfected the N terminal Yeast region, and was then able to flatten the Human hydrophobic extrema in the central region 300-680, producing a strong Human level hydrophobic level triplet in the center of the molecular motor for $W = W^* = 25$. How critical is this choice of W? The rms deviations from average for the three peaks is 0.3 for $W^* = 25$, increasing to ~ 2 for $W = 23, 27$ and 29, and doubling again for $W = 21, 31$. One can compare the $C^*$ correlations of BLAST aligned Human and Yeast profiles in Figs. 2 and 3 with BLAST values of Identity $I^* = 42\%$, and Positive $P^* = 59\%$. The profile values are ~ $C^*$ ~ 42%, so there are fundamental non-globular changes in Myosin-1 from Yeast to Human. Note that the central region is critical for correlating collective N- and C-terminal interactions, so it is natural that evolution improved this region in eukaryotes.

The discussion in [1] of actin evolution neglected the importance of actin/myosin complexes. There are two actin sequences, cytoplasmic (P60709) and skeletal muscle (P68133); only the latter was analyzed in [1]. According to BLAST, the two actin sequences are 96% positives. When we plot the MZ variance ratios $V_r$, we see a striking difference (Fig. 3): a minimum for skeletal muscle actin, but not for cytoplasmic actin. It is easy to see from this figure why conventional phylogenetics ($W = 1$) is unable to recognize Darwinian evolution [1].

Among the simplest early species are Fruit Fly and Round Worm. Here the BLAST values compared to Human are similar: Identity $I^*$ ~ 58%, and Positive $P^*$ ~ 72%. The surprise is that $C^*$ for both proteins' MZ $W = 25$ profiles has soared to 82%. This means that globular reshaping has dominated eukaryotic multicell myosin evolution. Note that the Human central level hydrophobic triad is no longer level in the more primitive species.

Fruit Fly and Round Worm sensitive middle regions are compared to Human in Figs. 5 and 6. The "floor" of level sets of hydrophilic minima shown in Figs. 2 and 3 for Human and Yeast is disturbed. Especially striking are the increased depths of the hydrophilic minima near 380, near the center of the molecular motor. The large scale motor motion can use a central mode which bends the N-terminal against the C-terminal, so we are not surprised to see a deep hydrophilic central hinge. The Human and Yeast central extrema of Ψ(aa,25) are 122 and 123, while Fruit Fly (118) and Round Worm (115) are more hydrophilic.

The broad evolution of myosin-1 discussed here shows large differences. Can scaling also discuss small differences, such as the difference between human myosin-1 (Uniprot 12882) and myosin-2 (UniprotQ9UKX2)? Here BLAST gives 94% identities and 97% positives. The correlation of Ψ(aa,25) for these two myosins is also 97%, very similar to BLAST positives. In addition, Fig. 6 shows that the differences are concentrated in regions related to actin binding. The changes occur in two places. First, there is a small leveling of My-1 hydrophobic extrema near the motor center, relative to My-2. Second, the heights of the hydrophilic and hydrophobic extrema near 600 are reversed. The actin binding regions are centered near 670 and 770. The reversal of extrema near 600 would presumably alter transition state kinetics prior to actin binding. Such symmetries are consistent with proximity to a functional critical point [12]. These extrema can affect the transition states preceding actin binding.

An interesting feature of Fig. 6 is that My-1 appears to be a subtle refinement of My-2, especially in the region of the motor center. My-2 is commonly found in muscle cells, while it has recently been reported that My-1 controls trafficking at the Golgi complex in a human cell line [15]. MYO1C stabilizes actin and facilitates the arrival of transport carriers at the Golgi complex. MYO1C is supposed to stabilize a scaffold around different intracellular compartments that allows better docking of cargo. The details discussed in the Fig. 6 caption involve level sets and network synchronization (see below).

As mentioned in the Introduction, one can use either first-order scaling with the KD scale [9] to treat functions associated with a few short-range strong interactions, or second-order scaling (MZ scale [11,12]) for functions involving many long-range weak interactions. The latter often evolve towards Ψ(aa,W*) profiles that exhibit level sets of both hydrophobic and hydrophilic extrema. We have seen many examples of level extrema with the MZ scale in Figs. 3, 5-7.

Many interesting things happen when we analyze Mys-1 human profiles with the KD scale and compare the results to those with the MZ scale (Fig. 7). Given that the correlation between the two scales is only 0.85 [12], whereas that for human Mys-1 with Mys-2 is 0.97, it would seem unlikely that the level extrema found with the MZ scale would persist to the KD scale. However, thermodynamically the distinction between first- and second-order transitions is definite, and we have found mixed first- and second-order effects in earlier studies of many proteins. In fact in Fig. 7 the two deepest hydrophilic minima, which were level with the MZ scale, are equally level with the KD scale! However, the hydrophobic maxima are drastically different. Below we will discuss the possibility that the minima are level to optimize Golgi interactions [15], while the maxima reflect first-order effects in actin binding. An enlargement of the actin binding region is shown in Fig. 8.

**Discussion**

One would expect to find large differences between multicell Human and single cell Yeast myocin, if only because the BLAST similarities are low. Fission Yeast has established itself as having the largest number of conserved elements in vertebrate genomes, while Humans have the smallest number [16]. Even so, finding a functional difference between Human and Yeast actomyosin complexes was difficult. It was finally solved by observing compressive cytokinetic rings, essential for separating daughter cells during division, oriented in single focal planes of micro-cavities [17,18]. In mammalian cells there was a direct transition from a homogeneous ring distribution to a periodic pattern of myosin clusters at the onset of constriction. In contrast, in fission yeast, myosin clusters rotate prior to and during constriction, explained by differences in the respective stresses.

Can these differences be connected to sequence differences at the molecular level? The N-terminal regions of Yeast show level extrema, while the Human do not. The central regions of Human and Yeast myosin show qualitatively different patterns in Fig. 2, with a strong stabilizing Human hydrophobic triplet spanning ~ 100 sites in the center of the molecular motor. This suggests that initially evolution favored a more active N-terminal in Yeast, but the dominant activity switched to the central region in Human. Yeast shows instead a broad and strong amphiphilic central region, spanning ~ 300 sites, where it can direct both terminals. This central region could support rotations driven by sliding water density waves. The latter could be coupled to polar and bipolar myosin ends [17, 18]. Here it is technically striking that Human triplet leveling is very accurate at $W = 25$. Differences between Human and Yeast myosin domains have been observed in many experiments [19]. A detailed analysis of these differences lies outside the scope of this article, but the general conclusion, that tail or lever domains are exchangeable, but motor domains are not [19], is consistent with the differences in motor domains identified here.

The evolutionary differences discussed here can be compared to analyses of myosin motor function involving highly conserved charged/polar residues [20]. These can also lead to long-range coupling between ATP-binding and lever-arm regions. As shown here, the details of this conserved coupling are modified by evolution, and the variable residues are refined hydrodynamically. The evolutionary refinement is not accessible to W = 1 BLAST and phylogenetics, but it is clear from the curvatures of large-scale (W ≈ 25) wave extremes. Overall the results discussed here are possible only because proteins evolve as continually improving self-organized networks approaching criticality [10,22,23].

The broad evolution of myosin-1 discussed here shows large differences (Figs. 1-6). As shown in Fig. 6 and its caption, there are subtle hydropathic differences between human Mys-1 and Mys-2. These small differences are consistent with proximity to a functional critical point [23]. The tilt of the two hydrophilic minima discussed in Fig. 6 for human Mys-2 is similar to that shown for fruit fly and worm Mys-1 (Figs. 5 and 6). Thus one could say that Mys-1 early in evolution is similar to Mys-2. It is only recently that evolution has optimized Mys-1 to function better as the Golgi complex traffic controller [15].

Because the differences are so small, we did not expect to see experimental data consistent our model. Thus the results of [15] are a pleasant surprise, and indicative of ever-increasing sophistication in the use of multiple experimental tools in cell biology. At the same time, the relelvance of thermodynamic descriptions of phase transitions [23] and general concepts of network synchronization [24,25] explain how such subtle and accurate differences in Ψ(aa,W*) can describe the evolution of the functional differences between Mys-1 and Mys-2.

One might have expected symmetry between the hydrophilic and hydrophobic extrema, in the sense that if the KD scale failed to show leveling of inner hydrophobic core extrema, it would also fail to show leveling of outer hydrophilic edge extrema, but this expected symmetry is broken (Figs. 7 and 8). At present core and edge interactions are being addressed by molecular dynamics simulations, but only for small proteins (~ 35 amino acids) [26,27]. In these studies protein folding pathways involve cascades of structural transitions from nearly unbound to observably bound state, and similar cascades are expected in the binding of Mys to actin.

How would Mys bind to actin? Binding involves a hydrophobic first-order transition, so one should look at the KD profile (Fig. 6). One might expect that near the start of the cascade, the Mys motor region would orient itself so that its center faced and hydrophobically approached the actin polymer chain. This is consistent with the triad of strong hydrophobic KD peaks near 500, with their nearly symmetric triangular positions, suitable for large-scale orientation. The cascade follows, sliding towards the region containing the two ground state binding regions, shown in more detail in Fig. 7. Note that the bidning regions are shifted towards the center of Mys (including the tail), which should be better balanced, While there is a small shift of the hydrophilic profile level minima from MZ to KD scales in Fig. 6, examination of Fig. 7 shows a much larger hydrophobic shift persists in the indicated binding regions.

How accurate are the MZ profiles for Myosin-1? According to BLAST, the chicken sequence P02565 registers 96% positives with the human sequence P12882. The important central regions are compared in Fig. 9. The two hydrophilic extrema are both level, and the human extrema are both 1.5 (0.5)% more hydrophilic than the chicken extrema. This accuracy of 0.5% can be compared to the BLAST positives difference of 4%. Comparison with Fig. 6 in the hydrophobic extrema shows that Myosin-2 human is similar to Myosin-1 chicken, and thus is less evolved than Myosin-1 human.

**Conclusions**

The success of the present method, with proteins containing thousands of amino acids, can be traced to at least two factors. First, its mathematical basis in topology: the in and out regions of proteins are accurately described for second-order transitions because the 20 fractals discovered in [10] were based on studies of solvent occlusion in > 5000 membrane-sized protein segments from recent high-resolution data. While this discovery has come as a surprise to nearly everyone (including the author), it is now confirmed by many studies of protein evolution [12], where excellent results, including synchronized level extrema, were obtained with the MZ scale. This scale is universally successful in treating long-range interactions. The second factor is the effectiveness of the choice W* = 25, which is a compromise between the coiled-coiled repeats, and the membrane interactions of the motor.

It is interesting to compare the detailed results of the present analysis to the broader impressions accessible to phylogenetic analysis [28]. There it was concluded that myosins-1 represent an evolutionarily ancient class of myosins that existed prior to the divergence of yeast. We have used this observation to examine how Darwinian evolution has refined the synchronization of myosin-1 motor domain motions from yeast through fruit fly and worm to human. The subtle Darwinian evolutionary differences identified here have achieved dramatic importance in the differences between Coronaviruses 2003 and 2019

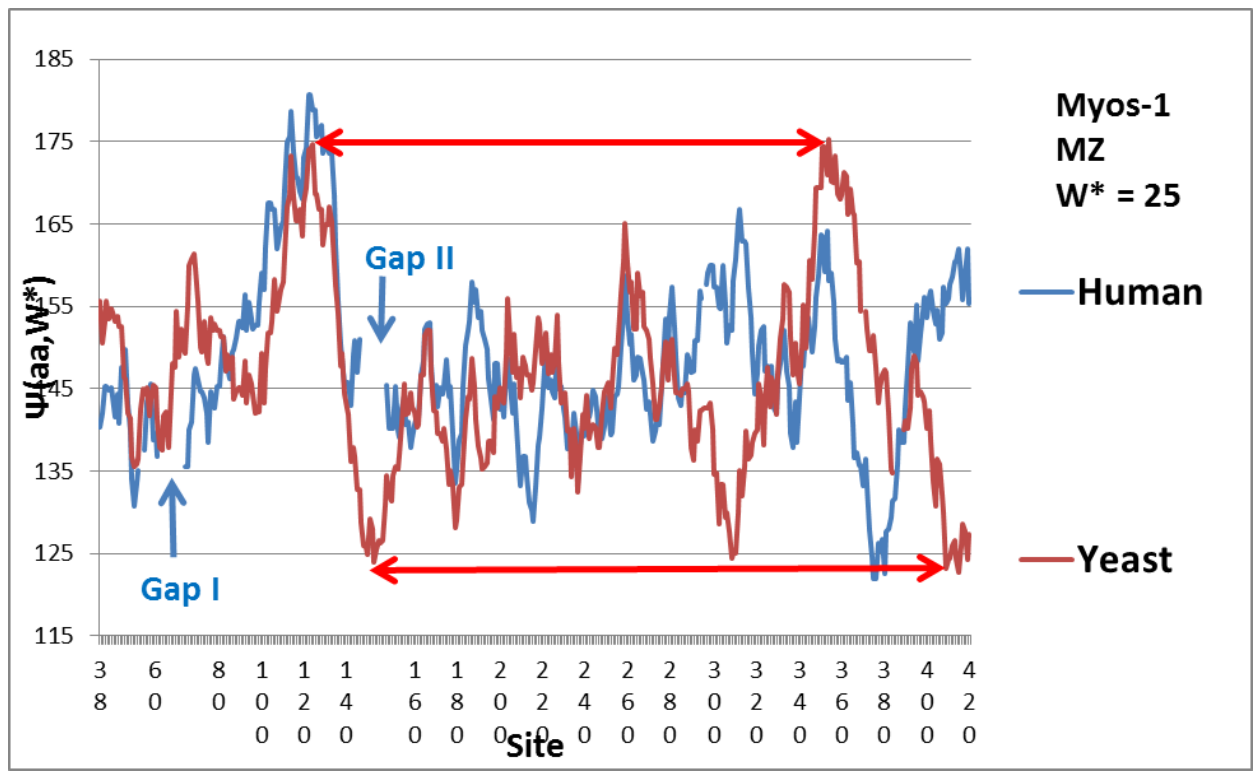

Fig. 1. Comparison of N - terminal Ψ(aa,W*) profiles (MZ scale) of Human and Yeast Myosin-1. BLAST alignment shows two 12 aa gaps in the Human sequence P12882 compared to the Yeast sequence P08964, as marked. Two major hydrophobic peaks and three major hydrophic valleys are also level in the Yeast profile, but not in the Human profile. Human site numbering.

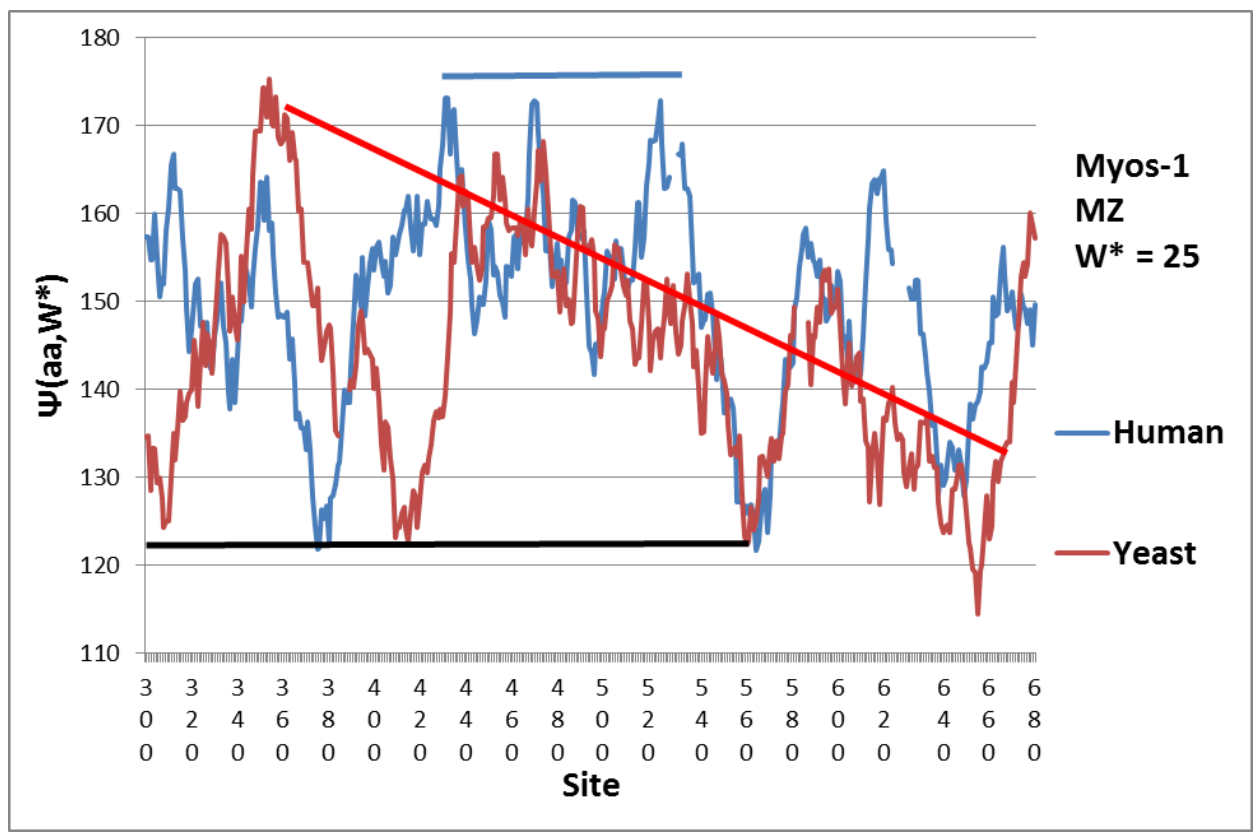

Fig. 2. Whereas in Fig. 1, in a wide Myosin-1 N-terminal region, Yeast showed more level sets, in the central region 300-680 shown here the Human hydrophobic extrema are more level. The Yeast extrema show a strongly amphiphilic (cascade) trend, while the Human hydrophobic extrema are flatter, with three level large hydrophobic extrema dominating the 430-550 region. Note that the hydrophilic floor remains for both cases at the same level as in Fig. 1. The larger actin-binding region 659-681 (Uniprot) lies at the end of the Yeast amphiphilic region (red line online).

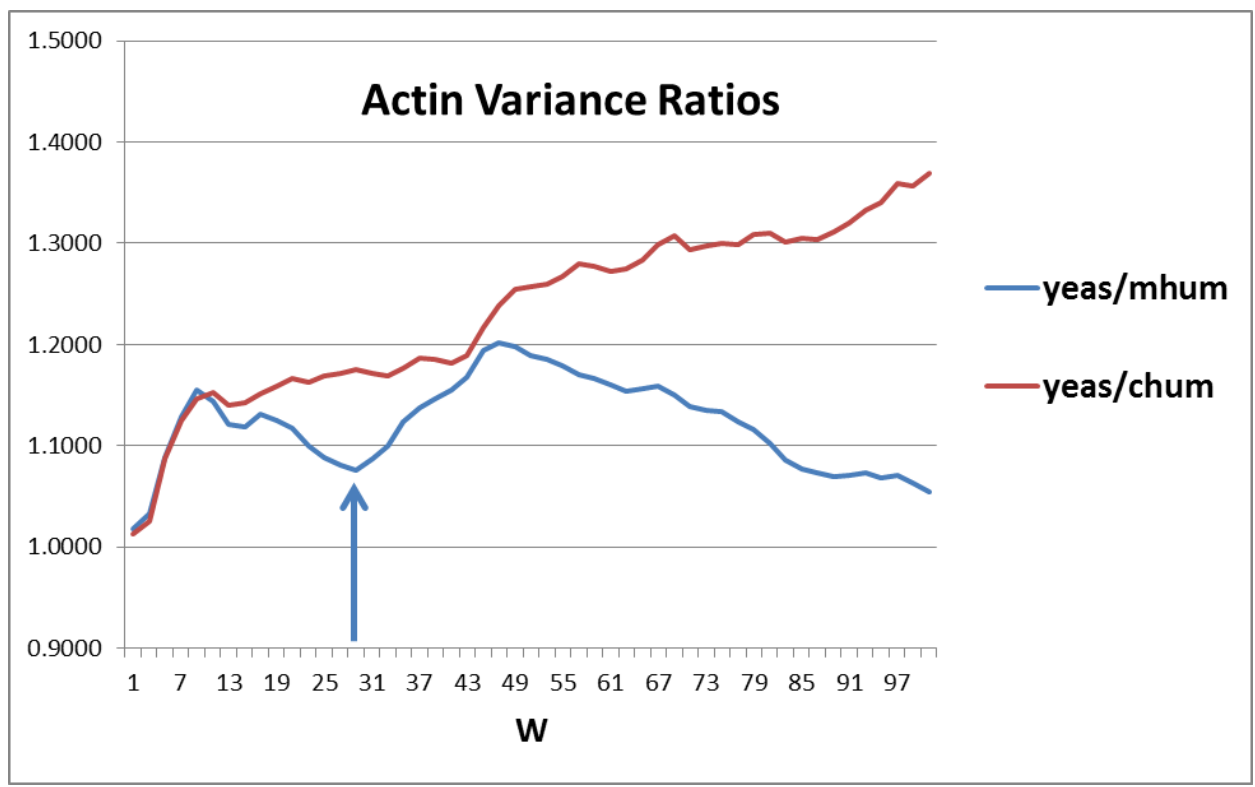

Fig. 3. Evolution of actin fission yeast (P60010) to actin (skeletal Muscle, P68133) and (Cytoplasmic, P60709). The two $V_r$ ratios are nearly identical up to W = 9 ( the lower limit of the MZ power-law range). For larger W (long-range interactions), there are large differences: a minimum is observed for skeletal Muscle actin at W = 29, but not for cytoplasmic actin. This minimum resonates with actin-myosin skeletal muscle complexes containing 28 amino acid repeat leaders. Note that this result has been obtained using no adjustable parameters. The divergence of the two $V_r$ ratios above W = 50 shows that skeletan muscle actin is reverting to yeast actin overall, while cytoplasmic actin's overall profile is narrowing and becoming less sensitive to water density fluctuations.

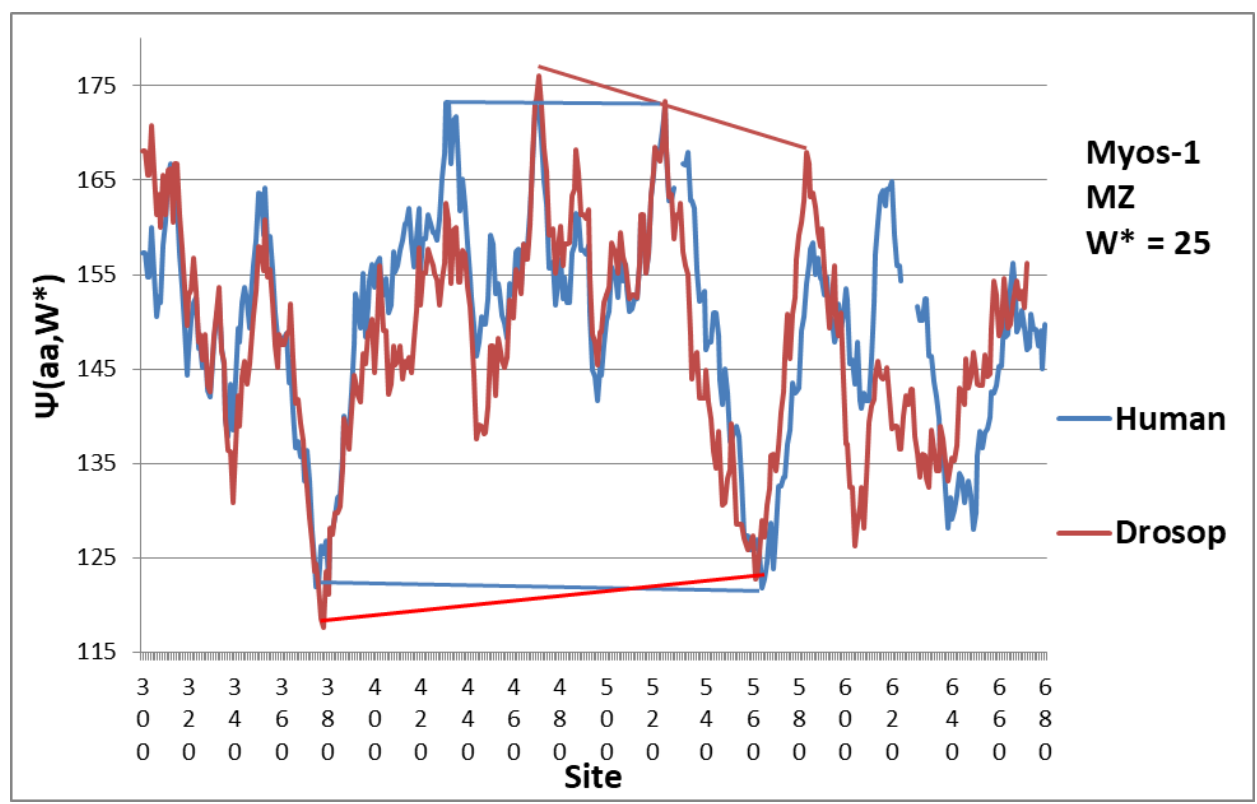

Fig. 4. The central region profiles compared for Myosin-1 Human and Fruit Fly. There are significant differences in the three highest hydrophobic peaks. The central trilevel hydrophobic peaks both linear, but in Human they are level, and they are tilted in Fruit Fly. A similar pattern was found [14] for Human and Slime Mold in the ubiquitin-activating enzyme Uba1 (E1). Note also the tilting of the two deepest hydrophilic minima in fruit fly.

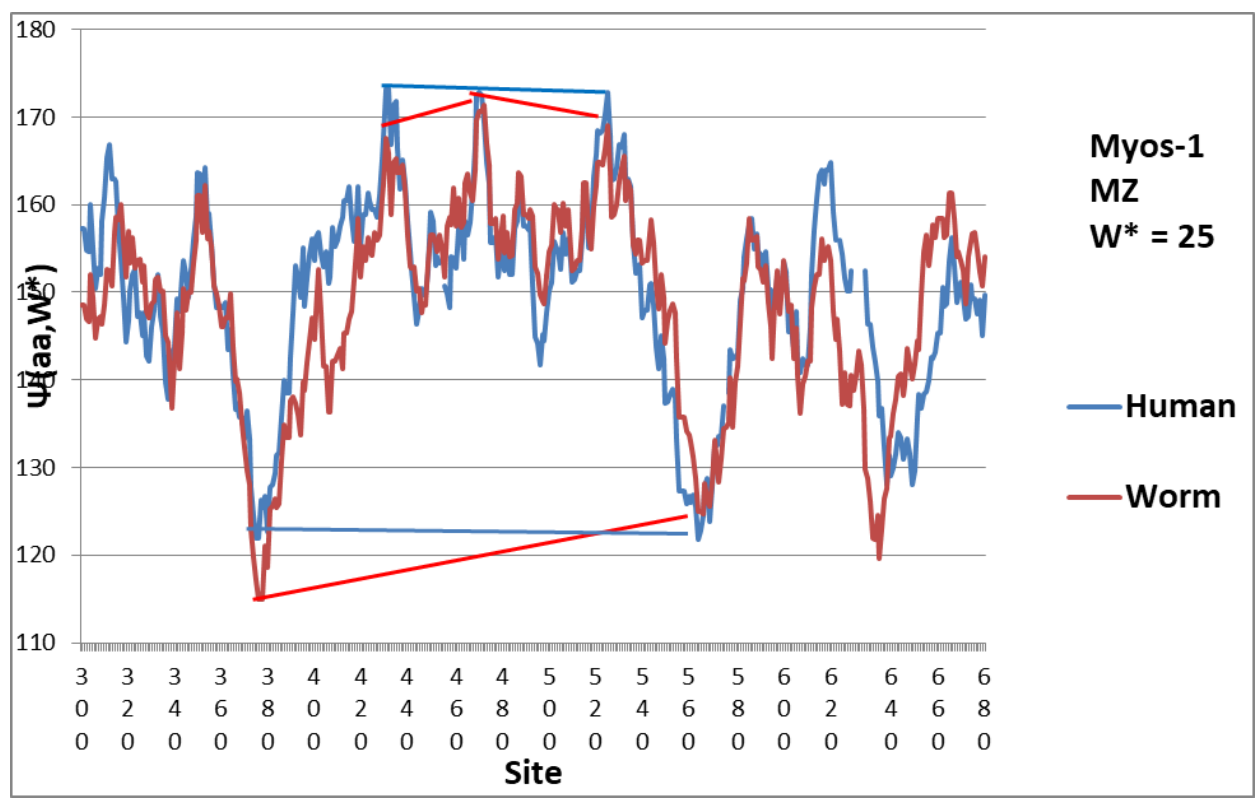

Fig. 5. The central region profiles compared for Human and Round Worm Myosin-1. As in Fig. 4, the differences from Human are small but significant. Here the central trilevel hydrophobic peaks of Worm are still less level than Human, but are more level than Fruit Fly. Note that the tilt of the deep minima near 370 and 570 is similar to that seen not only in fruit fly (Fig. 4), but also in human Mys-2 (Fig. 6).

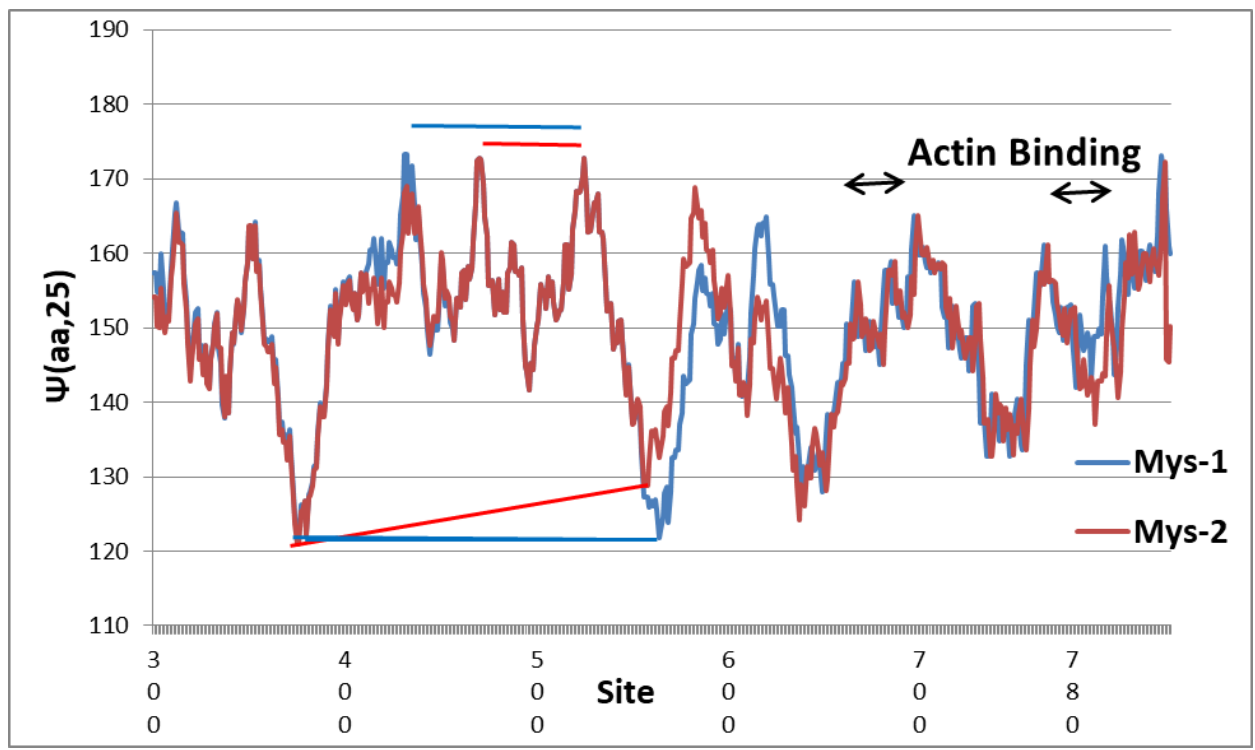

Fig. 6. The details of the small differences between human Mys-1 and Mys-2 show surprising symmetries, consistent with the differences in function discussed in the text and in [14]. To display better these differences only the 300-800 region is shown. The three central peaks of Mys-1 are at Ψ(aa,W*) = 173, as are the second and third peaks of Mys-2. The first peak of Mys-2 is slightly lower, at 169. The two hydrophilic extrema of Mys-1 are level at 122, whereas the similar extrema of Mys-2 are tilted at 121 and 129. The tilt of the human Mys-2 minima is similar to that seen in Mys-1 for worm (Fig. 5) and fruit fly (Fig. 4). Note also the reversal of the relative heights of the four extrema near 600.

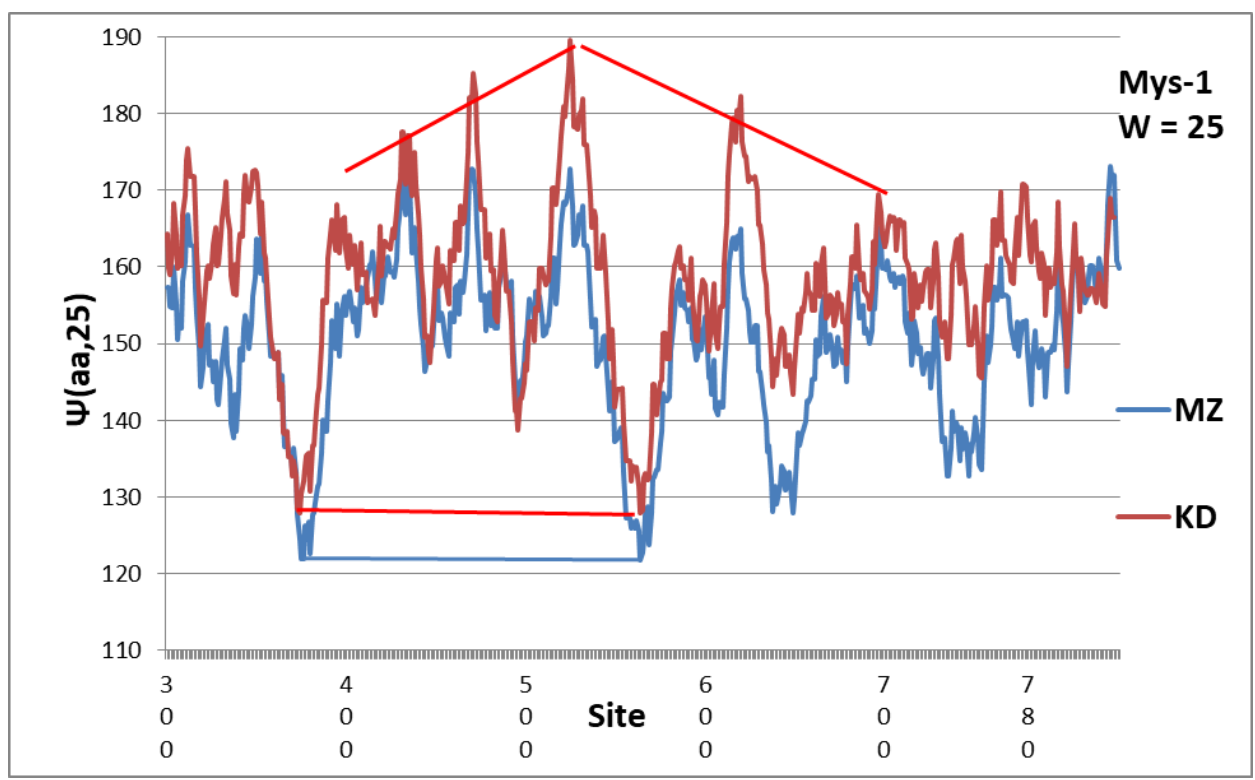

Fig. 7. The KD profile retains the hydrophilic minima of Mys-1 and they are still level to <1% difference, which is surprising (see text). The three hydrophobic maxima which were level to <1% with the MZ scale are no longer level, but exhibit a triangular maximum.

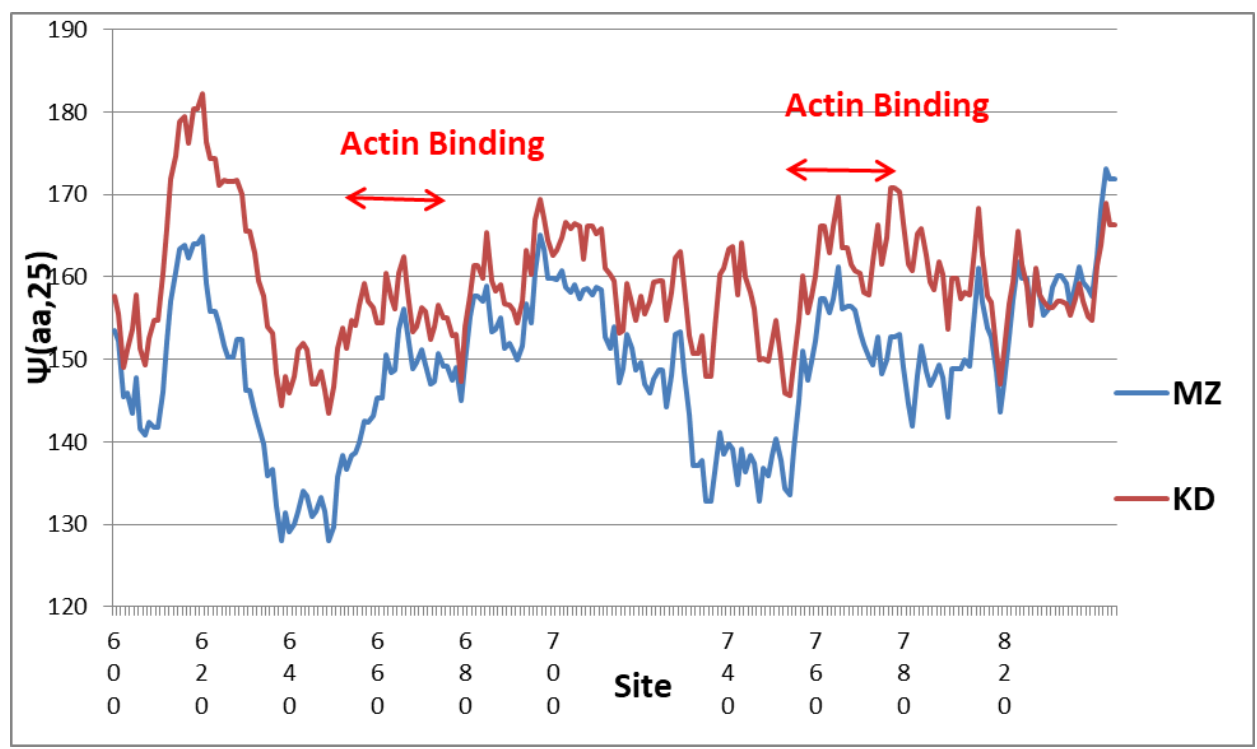

Fig. 8. At first glance, the human Mys-1 actin binding regions with the KD scale resemble those with the MZ scale, but closer inspection shows that the KD scale actin binding regions are more hydrophobic.

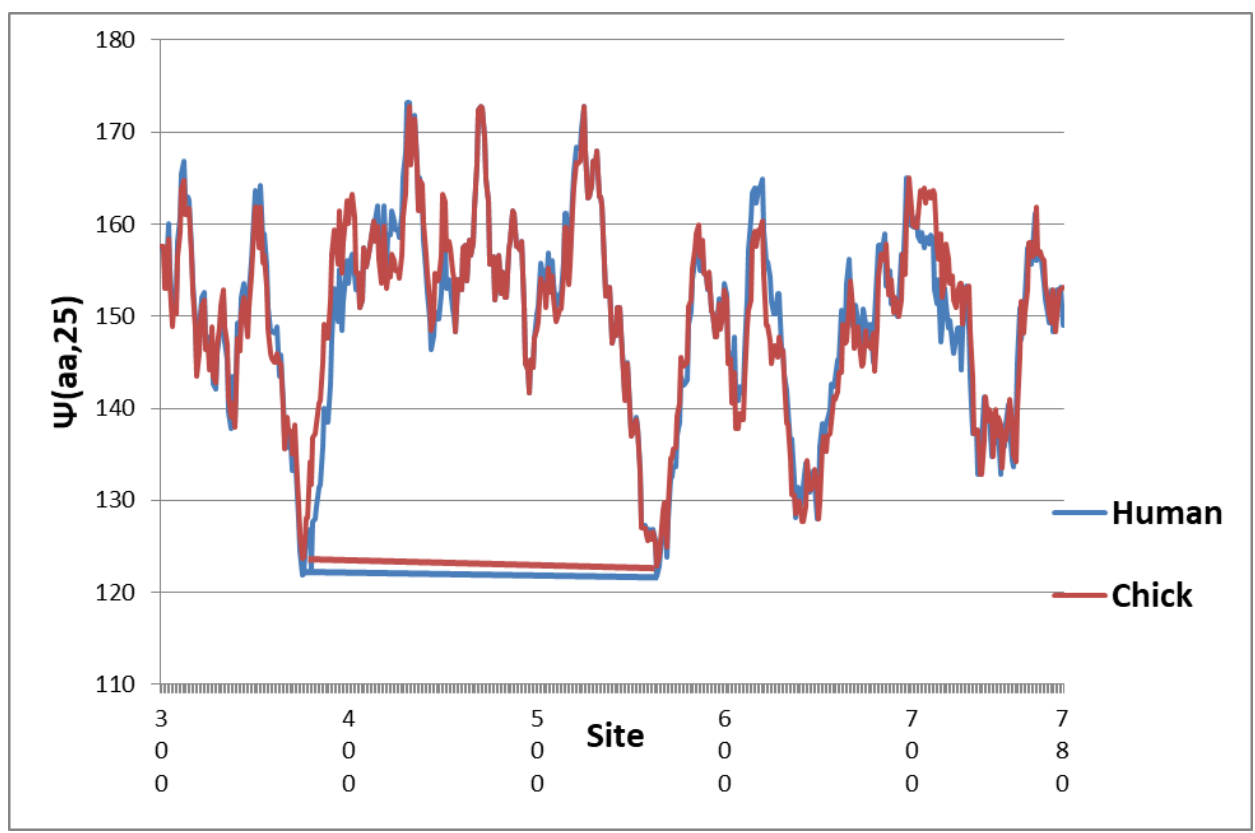

Fig. 9. Because chicken and human Myosin-1 are so similar, their sequences can be used to test the accuracy of the MZ scale. The two hydrophilic extrema are both level, and the human extrema are both 1.5 (0.5)% more hydrophilic than the chicken extrema. This accuracy of 0.5% can be compared to the BLAST positives difference of 4%. Comparison with Fig. 6 in the hydrophobic extrema shows that Myosin-2 is similar to Myosin-1 chicken, and thus is less evolved than Myosin-1.